\documentclass[12pt]{article}
\usepackage{amsmath,amsfonts}
\usepackage{graphicx}
\input epsf

\makeatletter\@addtoreset{equation}{section}\makeatother

\def\IC{\mathbb{C}}

\def\IZ{{\mathbb{Z}}}
\def\IR{{\mathbb{R}}}
\def\IP{\mathbb{P}}

\def\CN {{\cal N}}

\newcommand {\apgt} {\ {\raise-.5ex\hbox{$\buildrel>\over\sim$}}\ }
\newcommand {\aplt} {\ {\raise-.5ex\hbox{$\buildrel<\over\sim$}}\ }

\makeatletter\@addtoreset{equation}{section}\makeatother

\newcommand{\Tr}{{\rm Tr\,}}

\renewcommand{\title}[1]{\vbox{\center\LARGE{#1}}\vspace{5mm}}
\renewcommand{\author}[1]{\vbox{\center#1}\vspace{5mm}}
\newcommand{\address}[1]{\vbox{\center\em#1}}
\newcommand{\email}[1]{\vbox{\center\tt#1}\vspace{5mm}}

\begin{document}

\unitlength = .8mm

\begin{center}
\hfill \\
\hfill \\
\vskip 1cm

\title{Notes on adding D6 branes wrapping $\IR\IP^3$ in AdS$_4 \times \IC\IP^3$}

\author{Davide Gaiotto$^{1,a}$ and Daniel Louis Jafferis$^{2,b}$}

\address{$^1$School of Natural Sciences, Institute for Advanced Study, \\Princeton, NJ 08540, USA \\$^2$NHETC and Department of Physics and Astronomy, Rutgers University,\\ Piscataway, NJ 08855, USA}

\email{$^a$dgaiotto@ias.edu \\ $^b$jafferis@physics.rutgers.edu}

\end{center}

\abstract{We deform the ${\cal N}=6$ Chern Simons theory by adding
extra matter hypermultiplets in a fundamental representation of
one or both gauge groups. We compute the quantum corrected moduli
space. We verify that the holographic dual of the modified theory
consists of the usual AdS$_4 \times \IC\IP^3$ background in
presence of AdS$_4$ filling D6 branes which wrap $\IR\IP^3 \subset
\IC\IP^3$. We extend the correspondence to a similar modification
of more general known ${\cal N}=3$ dual pairs}



\section{Introduction}

${\cal N}=3$ CSM theories in three dimensions have a powerful
property: they have a classically and conformally invariant action
for any choice of matter content, with no marginal deformations. In
this note we consider a modification of the ${\cal N}=6$ ABJM theory
\cite{ABJM}, where extra matter is added, in the fundamental
representations $(N,1)$ or $(1,N)$ of the $U(N) \times U(N)$ gauge
group. This breaks the $\CN = 6$ supersymmetry to $\CN = 3$, but
preserves conformal invariance. We relate the resulting theory to
Type IIA string theory on $AdS_4 \times \IC\IP^3$ in presence of
 AdS$_4$ filling D6 branes which wrap $\IR\IP^3 \subset
\IC\IP^3$. In general there can be $m_1$ fundamentals on the first
node, and $m_2$ on the second. This choice corresponds in the IIA
picture to the choice of the $\IZ_2$ Wilson line on the $m_1+m_2$
D6, living in $\pi_1(\IR\IP^3)$. The IIA setup lifts to a purely
geometric background of M-theory,  $AdS_4 \times M_7$ for certain
$3$-Sasakian manifolds $M_7$. In M-theory, the choice of Wilson line
lifts to a flat topologically nontrivial C-field.

There is a branch of the moduli space of the Chern-Simons-matter
theory that corresponds to M2 branes probing this geometry. The
classical moduli space in that branch receives a quantum
correction. The 1-loop exact quantum correction to the metric on
the hyperk\"{a}hler moduli space was found by direct calculation
in \cite{jafferis-yin}. Here we construct the chiral ring. It
looks very similar to the chiral ring of the uncorrected theory,
except that the conformal dimension of the monopole operators that
appear in certain chiral primaries becomes non-zero due to the
presence of the fundamental matter. This beautifully matches the
chiral ring of the expected moduli space. In the special case
where the Chern Simons level is $1$, and a single D6 brane is
added, we can see the $SU(3)$ isometry of the resulting
$N^{0,1,0}$ 3-Sasakian near horizon geometry.

There is another branch to the moduli space, in which the $N$ D2
branes become dissolved in the $m$ D6 branes. This is expected to
give the moduli space of $N$ instantons of rank $m$ on the cone over
$\IR\IP^3$, $\IC^2/\IZ_2$. The Chern-Simons levels do not enter the
analysis of this branch of the moduli space, and the result is
identical to the Higgs branch of a Yang-Mills theory with the same
matter content. In fact, this branch is characterized by the fact
that the fundamentals have nonzero VEVs. This implies that the
moment maps must all be vanishing due to the F-term and D-term
equations. Then this branch of the moduli space of this 2+1 CSM
quiver theory is exactly the moduli space of the same quiver
interpreted as a 3+1 gauge theory. In fact the quiver is precisely
the ADHM quiver for charge $N$ instantons of rank $m_1 + m_2$ on
$\IC^2/\IZ_2$. This agrees beautifully with the expected result.

One can study the same question for the more general $\CN=3$ quiver
CSM theories with $n$ nodes of \cite{imamura-kimura,
jafferis-tomasiello}. The branch in which the D2 branes are
dissolved into the D6 branes again leads exactly to the ADHM quiver,
this time for $\IC^2/\IZ_n$. Indeed in this case the D6 branes are
wrapping an $S^3/\IZ_n$ homologically trivial 3-cycle.

After this draft was completed, we received
\cite{Hohenegger:2009as}, which has significant overlap with this
work, but reaches different conclusions regarding the precise match
between the number of $D6$ branes and the number of fundamental
fields $m_1,m_2$.

\section{Adding fundamentals to Chern-Simons-matter theories}

Consider the introduction of fundamental hypermultiplets in the
$\CN=6$ theory of \cite{ABJM}. This cannot be done in a manner
preserving more than $\CN=3$ supersymmetry. The Lagrangian for a
$\CN=3$ theory, with unbroken $SO(3)$ R-symmetry, with two
bifundamental hypermultiplets and any number of fundamental
hypermultiplets is uniquely determined. We will denote the
bifundamental chiral fields $A^i$, $B_i$ for $i=1,2$, $m_1$
fundamentals of the first gauge group, $a^t$, $b_t$, and $m_2$
fundamentals of the second group, $c^s$, $d_s$.

The $N=2$ superpotential for the $\CN=3$ theory is
\begin{equation}
\frac{4 \pi}{k_1}(B_i A^i  + b_t a^t )^2 - \frac{4 \pi}{k_2}(- A^i B_i - c^s d_s)^2
\end{equation}
We see that even if $k_1 + k_2=0$, the coupling to the fundamental
matter reduces the flavor symmetry to an $SO(3)_F$. This
illustrates clearly the fact that that the $\CN=3$ supersymmetry
is not enhanced, as the original $SO(6)$ R-symmetry is reduced to
$SO(3)_R \times SO(3)_F$. Although in the following we will keep
$k_1 + k_2=0$, it would be very natural to relax this constraint,
and introduce a Roman mass in the holographic dual, as in
\cite{Gaiotto:2009mv}. Along the same lines, we can add
fundamental matter to a more general family $\CN=3$ CSM theories
with known holographic duals. These theories were introduced in
\cite{jafferis-tomasiello}. They are the unique $\CN=3$ quiver
theories with unitary gauge groups organized in a necklace, with a
single bifundamental hypermultiplet between adjacent nodes. If we
add fundamental matter while preserving $\CN=3$ SUSY we get a
superpotential
\begin{equation}
\sum_i \frac{4 \pi}{k_i}(B^{(i+1)} A^{(i+1)}- A^{(i)} B^{(i)} + b^{(i)}_t a^{(i),t} )^2
\end{equation}
The resulting $\CN=3$ Chern-Simons-matter theory with fundamentals
will still be conformally invariant, since there are no marginal
or relevant operators which preserve the $\CN=3$ supersymmetry and
$SO(3)$ R-symmetry, just as in the case without fundamentals
\cite{gaiotto-yin}.

\section{Introducing D6 branes in AdS$_4 \times M_6$}

\subsection{D5 branes in the IIB configuration}

We begin with IIB theory on a circle, with axio-dilaton $\tau =
\frac{i}{g_s} + \chi$, and consider $N$ D3 branes along directions
$0123$ (where $x^3$ is the circle direction), and various D5 and
$(1,p_i)$ fivebranes. The $(1,p_i)$ fivebrane, $i=1,..., n$, is
extended along $012[37]_{\theta_i} [48]_{\theta_i} [59]_{\theta_i}$,
where $\theta_i = \arg(\tau) - \arg(p_i + \tau)$. The $m$ D5 branes
are extended along the 3-plane with $\theta = \arg(\tau)$. This
configuration preserves $\CN=3$ supersymmetry \cite{kitao-ohta-ohta,
bergman-hanany-karch-kol}. In the simplest example, D5 branes are
added to the configuration that engineers the $\CN=6$
Chern-Simons-matter theory \cite{ABJM}, with one NS5 brane and one
$(1,k)$ fivebrane.

The T-dual and lift to M-theory of such a configuration of $(p,q)$
fivebranes was determined by \cite{GGPT}. In general, a Lagrangian
description of the effective 2+1 field theory on $N$ D3 branes
stretched between a $(p,q)$ fivebrane and a $(p', q')$ fivebrane is
not known. When all the fivebranes have a single unit of NS5 charge,
the effective field on the D3 branes flows to a 2+1 conformal field
given by an $\CN=3$ quiver Chern-Simons-matter theory
\cite{imamura-kimura, jafferis-tomasiello}. The Chern-Simons level
on the D3 branes stretched between the successive $(1,p_i)$ and
$(1,p_{i+1})$ fivebranes is given by $k_i = p_{i+1}-p_i$. We now
consider introducing some D5 branes as well.

The D5 branes may be distributed along the the $x^3$ circle
between any pair of $(1,p_i)$ fivebranes, so we partition $m =
\sum_{i=1}^n m_i$. There will be $m_i$ fundamental hypermultiplets
attached to that node in the quiver, arising from the D3-D5
bifundamental strings. Note that in this configuration there are 4
Neumann-Dirichlet directions, thus one obtains precisely two
chiral multiplets that make up the hypermultiplet.

\subsection{M-theory lift} \label{lift}

This IIB configuration can be T-dualized and lifted to M-theory,
following \cite{GGPT}. Applying T-duality to the $x^3$ circle
takes the D3 branes to D2 branes, now living in a seven
dimensional transverse geometry. The $(1,p_i)$  fivebranes become
Taub-NUT with D6 charge dissolved into $F_2$ flux. The D5 branes
naturally become D6 branes in this geometry. The metric is much
simpler after lifting to M-theory, so we will postpone the details
of the IIA description until later.

This lifts to pure geometry in M-theory, with M2 branes probing an
eight dimensional hyperkahler transverse space. As shown in
\cite{GGPT}, the metric on this $T^2$ fibration over a base
$\IR^6$ is given in terms of a two by two matrix of harmonic
functions as \begin{equation} \begin{split}
  ds^2 = & U_{ij} d \vec x^i \cdot d \vec x^j + U^{ij} ( d \varphi_i + A_i ) ( d \varphi_j  +
  A_j) ,
 \\
 A_i = & d \vec x^j \cdot \vec \omega_{ji }  = dx^j_a \omega^a_{ji} ~,~~~
 ~~ ~~~ \partial_{ x_a^j} \omega^{b}_{k i } - \partial_{x_b^k} \omega^a_{ji} =
  \epsilon^{ab c} \partial_{x^j_c} U_{ki}~,
  \end{split}
 \end{equation}
 where $U^{ij}$ is the inverse of the matrix $U_{ij}$. The matrix $U$ obeys linear
  equations that follow
 from this ansatz.
 The metric of a single Kaluza-Klein monopole times
 $\IR^3\times S^1$ that arises from a single NS5 brane
 can be written in this form as a configuration with
 \begin{equation}
 U = U_{\infty} +   \left(\begin{array}{cc} h_1 & 0 \\ 0 & 0 \end{array} \right) ~,~~~~~~ h_1=  \frac{1}{2 |\vec x_1|}.
 \end{equation}
 The asymptotic value $U_\infty$ encodes the complex and Kahler
 parameters of the torus fiber at infinity, which are determined
 in terms of the original IIB coupling $\tau$, and the radius of
 the $x^3$ circle. The matrix of harmonic functions associated to
 a $(p,q)$ fivebrane can be obtained from the above $U$ by
 application of the appropriate $SL(2,\IZ)$ transformation via
 \begin{equation} U \mapsto g^\dag U g, \ \ \ \ \left(\begin{array}{c}
 {\vec x}_1 \\ {\vec x}_2 \end{array} \right) \mapsto g \left(\begin{array}{c}
 {\vec x}_1 \\ {\vec x}_2 \end{array} \right).\end{equation}

The linearity of the harmonic equation that $U$ satisfies implies
that the metric obtained by lifting $(1,q_i)$ fivebranes, and $m$
D5 branes is given by \begin{equation} U = U_{\infty} +
\sum_{i=1}^n \frac{1}{2 |\vec x_1 + q_i \vec x_2|} \left(\begin{array}{cc} 1 & q_i \\
q_i & q_i^2
\end{array} \right) + \frac{m}{2|\vec x_2|} \left(\begin{array}{cc} 0 & 0 \\ 0 & 1 \end{array} \right).
 \end{equation}

The low energy effective theory on the stack of M2 branes will be
determined by the local singularity at the origin in this geometry.
This conical hyperkahler 8-manifold was shown by
\cite{bielawski-dancer} to be a particular abelian hyperkahler
quotient. In particular, the pair of $U(1)$ isometries of the $T^2$
fiber are compatible with the hyperkahler structure, and one obtains
the hypertoric manifold $\mathbb{H}^{n+1}/// \bf{N}$, where $\bf{N}$
is the kernel of the map
\begin{equation}\label{eq:beta}
    \beta: {\rm U}(1)^{n+1}\to {\rm U}(1)^2, \qquad \beta= \left(\begin{array}{ccccc}
   1 & 1 & \ldots & 1 & 0\\
  p_1 & p_2 & \ldots &p_n& m
    \end{array}\right)\ .
\end{equation}
The hyperkahler quotient just described sets to zero the moment maps
for all $U(1)$'s in the kernel of the $\beta$, thus we see that the
coordinates on the base $\IR^6$ are exactly
\begin{equation}\begin{split} \mu_I^\alpha = \sum_{i=1}^n
\mu^\alpha_i, \\
\mu_{II}^\alpha = \sum_{i=1}^n p_i \mu^\alpha_i + m \mu^\alpha_{n+1}.
\end{split}\end{equation}

The way the lifted geometry is modified from that found in
\cite{jafferis-tomasiello} by the inclusion of the D5 branes can be
explained as follows. Consider the geometry obtained without D5
branes. If we erase the $D5$ brane, we have the hypertoric manifold
$\mathbb{H}^{n}/// \bf{\tilde N}$, where $\bf{\tilde N}$ is the
kernel of the map
\begin{equation}\label{eq:betap}
    \beta: {\rm U}(1)^{n}\to {\rm U}(1)^2, \qquad \beta= \left(\begin{array}{ccccc}
   1 & 1 & \ldots & 1 \\
  p_1 & p_2 & \ldots &p_n
    \end{array}\right)\ .
\end{equation}
The coordinates on the base $\IR^6$ are exactly
\begin{equation}\begin{split} \mu_I^\alpha = \sum_{i=1}^n
\mu^\alpha_i, \\
\mu_{II}^\alpha = \sum_{i=1}^n p_i \mu^\alpha_i.
\end{split}\end{equation}
In the IIA picture, before going to the near horizon limit, the $m$
D6 branes wrap the cycle defined by $\mu_{II}^\alpha = 0$. We will
better characterize this cycle in the next subsection. For now,
let's keep looking at the backreacted M-theory geometry.

We want to compare the ``chiral rings'' on the two manifolds, i.e.
the set of homogeneous, holomorphic functions on the hyperkahler
cones in a given complex structure. In a hypertoric manifold
$\mathbb{H}^{n+1}/// \bf{N}$, such functions are built out of
monomials of the elementary quaternionic coordinates on
$\mathbb{H}^{n+1}$, invariant under $\bf{N}$, modulo F-term
relations.

With respect to $\mathbb{H}^{n}/// \bf{\tilde N}$ we have an extra
quaternionic coordinate $(u,v)$. The set of F-term relations
\begin{equation}
\sum c_i u_i v_i + c u v =0,   \qquad \sum c_i=0 \qquad \sum c_i p_i + m c=0
\end{equation}
is the union of the F-term relations $c=0$ for $\mathbb{H}^{n}///
\bf{\tilde N}$ and an extra one, which can be used to eliminate $u
v$ from the monomials. We are led to look for functions of the
general form $u^d f$ (for $d\geq0$) or $v^{-d} f$ (for $d\leq0$).
Here $f$ is a function on $\mathbb{H}^{n}/// \bf{\tilde N}$, which
transforms with definite weights, $a_1$, $a_2$, under the pair of
hypertoric $U(1)$ isometries, $(\lambda_1, \lambda_2): f \mapsto
\lambda_1^{a_1} \lambda_2^{a_2} f$. Now we should require $u^d f$
($v^{-d} f$) to be invariant under all the $U(1)$'s except those in
the kernel of the new $\beta$.

Of course, $f$ is already invariant under the quotients of the
original geometry, so the only new requirement is that under a
transformation with weights $(\lambda_1, ..., \lambda_n,
\lambda_0)$, such that $\prod_{i=1}^n \lambda_i = 1$ and
$\lambda_0^m \prod_{i=1}^n \lambda_i^{p_i} = 1$, then \\ $u^d f
\mapsto \lambda_0^d (\prod_{i=1}^n \lambda_i^{p_i})^{a_2} u^d f$
must be invariant. Therefore we must have $$d = m a_2.$$ This says
that the functions on the new hyperkahler manifold are the same as
before, but dressed with an appropriate power of $u$. Thinking of
these as elements in the chiral ring, we see that the spectrum of
chiral operators is unchanged, but the conformal dimensions are
shifted, by precisely $\frac{1}{2} m$ times the baryon number of
the operator. The chiral ring relations are also modified, because
the product $u^{d_1} f_1 v^{d_2} f_2$ will have to be rewritten to
eliminate the factors of $u v$ with the F-term relation.

\subsection{D6 branes on $\IR\IP^3$ in $\IC\IP^3$}

Reducing to IIA on the $S^1$ associated to the baryonic current in
the field theory, we obtain $N$ D2 branes in a seven dimensional
transverse cone. The near horizon geometry is given by AdS$_4
\times \IC\IP^3$ in the $\CN=6$ case, and for more general quivers
the dilaton is varying in the internal six manifold.

Note that the D5 branes introduced above become D6 branes in the IIA reduction of the near horizon geometry. That is, the same circle
is identified as the M-theory circle in the GGPT geometry and the 3-Sasakian near horizon. Intuitively, this is because we are taking an
't Hooft limit for which the D5 charge goes to infinity while the NS5 charge remains fixed in the IIB configuration, hence the smallest cycle
will be the M-theory circle that shrinks due to the D5 charge.

More concretely, recall that the 3-Sasakian internal seven manifold
is the unit sphere in the singular hyperkahler quotient
$\mathbb{H}^{n+1}/// \bf{N}$. The 't Hooft limit involves scaling
all of the fivebrane charges by $(1,q_i) \mapsto (1,k q_i)$, hence
the kernel of $\beta$ contains a discrete subgroup $\mathbb{Z}_k$
acting by phase rotation on all of the $u_i$. This $\mathbb{Z}_k$
sits inside of a $U(1)$ isometry of the hyperkahler eight manifold,
and results in a parametrically small cycle in the near horizon
geometry. But this $U(1)$ is precisely the phase associated with the
coordinate $\mu_{II}^\alpha$ on the base of the GGPT torus
fibration. Thus we see that the cycle we reduce on to IIA is the
same as the M-theory circle in the lift of the original brane
configuration.

As we said above, the D6 branes are wrapping the locus
$\mu^\alpha_{II}=0$ in the seven dimensional cone. Notice that if we
set $\mu^\alpha_{II}=0$ and remove the corresponding $U(1)$ circle
reduce to IIA theory, we are really doing an hyperkahler quotient.
More precisely, the D6 branes are wrapped on the hyperkahler four
manifold given by the quotient $\mathbb{H}^n
///\bf{N}'$, where $\bf{N}'$ is the kernel of the map $$\beta': U(1)^n
\rightarrow U(1), \ \ \ \ \ \beta' = (1 \ 1 \ ... \ 1).$$ This is a
standard realization of the hyperkahler geometry $\IC^2/\IZ_n$. In
the near horizon limit, the cycle wrapped by the D6 branes is the
unit sphere in this space, namely $S^3/\IZ_n$. For $n=2$, this is
$\IR\IP^3$ in $\IC\IP^3$. It is preserved by a $SO(4) \in SU(4)$
subgroup of the isometries of $\IC\IP^3$, which coincides with the
symmetry group of our proposed SCFT.

Notice that $\pi_1(S^3/\IZ_n) =\IZ_n$, hence the D6 branes can carry
a discrete $\IZ_n$ Wilson line. This discrete parameter in the IIA
description is the remnant of the position of the D5 branes in the
IIB circle. They could sit in any of the $n$ intervals between
$(1,p_i)$ and $(1,p_{i+1})$ fivebranes. They would correspondingly
contribute a single fundamental hypermultiplet at the node of the
necklace quiver with Chern-Simons coupling $p_{i+1}-p_i$.

\subsection{Volumes and Free Energy}

Consider the M-theory lift of AdS$_4 \times \IP^3$ with $m$ D6
branes wrapping the $\IR\IP^3$ cycle. The internal tri-Sasakian
seven manifold is the unit sphere in the hypertoric eight manifold
$\mathbb{H}^3///U(1)$, where the $U(1)$ acts with charges $m, m, k$.
The volume of the seven manifold is given by \cite{lee-yee}
\begin{equation} Vol(M_7) = Vol(S^7)
\frac{m+2k}{2(m+k)^2},\end{equation} in terms of the volume of a
sphere with the same radius of curvature. In the AdS$_4 \times M_7$
near horizon limit, the total integral of $\ast G_4$ on $M_7$ is by
definition the total M2 brane charge of $N$. The effective four
dimensional supergravity solution corresponding to a black M2 brane
only depends on the Planck scale and the local value of the four
form field strength in the AdS$_4$.

The number of degrees of freedom at high temperatures is determined
from the AdS black hole, which is modified from the calculation of
M2 branes in flat space only by the change in the four dimensional
Plank scale, i.e. in the volume of the internal space.
We find that in area $V_2$ and at temperature $T$:
$$\beta F = -2^{7/2} 3^{-2} \pi^2  N^{3/2} \frac{(m+k)\sqrt{2}}{\sqrt{m+2k}} V_2 T^2 \sim \frac{N^2}{\sqrt{\lambda}} +
\frac{3 m N}{4} \sqrt{\lambda} + ...$$ Intriguingly, there is
actually an {\it enhancement} of the number of degrees of freedom in
the fundamentals by a factor of $\sqrt{\lambda}$, relative to $m$
weakly coupled $U(N)$ fundamentals! Of course, the total number of
degrees of freedom is still less than the counting of fields in a
free theory, $N^2 + m N$ in this case, since we are assuming $m \ll
k$ in the above expansion. It would be interesting to explore this
phenomenon further.

\section{Higgs branch moduli space}

The superpotential is given by \begin{equation} W = \frac{1}{k}
\Tr\left( B_i A^i + b_t a^t\right)^2 - \frac{1}{k} \Tr\left( -A^i
B_i - c^s d_s \right)^2.\end{equation} Define the hyperk\"ahler
moment maps \begin{equation}\begin{split} \mu_1^\alpha = \left\{B_i
A^i + b_t a^t, A_i^\dag A^i - B_i B^{i
\dag} + a_t^\dag a^t - b_t b^{t \dag} \right\} \\
\mu_2^\alpha = \left\{- A^i B_i - c^s d_s, B^{i \dag} B_i - A^i
A_i^\dag + d^{s \dag} d_s - c^s c_s^\dag \right\}.
\end{split}\end{equation} Then the bosonic potential vanishes if
and only if \begin{equation} \begin{split} A^i
\frac{\mu^\alpha_1}{k} = \frac{\mu^\alpha_2}{-k} A^i \qquad
 \frac{\mu^\alpha_1}{k} B_i = B_i \frac{\mu^\alpha_2}{-k}\\
a^t \frac{\mu^\alpha_1}{k} = 0\qquad \frac{\mu^\alpha_1}{k} b_t =
0\qquad
 \frac{\mu^\alpha_2}{-k} c^s = 0 \qquad
d_s \frac{\mu^\alpha_1}{k} = 0. \end{split}\end{equation} The branch
in which the D2 branes dissolve into the D6 branes is
$\mu_a^\alpha=0$. Only in that case can one turn on the
fundamentals. Now there is no Mukhi effect, since there are matter
fields turned on which are charged under the entire gauge group.
That is, the gauge symmetry is completely Higgsed. Therefore solving
the F-term equations, together with $\mu_a^{\IR} = 0$, gives exactly
the moduli space of the same quiver interpreted as a 3+1 Yang-Mills
quiver. But it is just the ADHM quiver describing $N$ instantons in
$\IC^2/\IZ_2$ with rank $m_1+m_2$.

More generally, the hyperkahler moduli space of a general ${\cal
N}=3$ theory is determined as follows. The  bosonic potential is
given by $$V = \sum_{\alpha=1}^3 \sum_a \left| (k^{-1})^{i j}
\mu_i^\alpha T^{a b}_j q_b^A \right|^2$$ where where $q_b^A$ are the
matter fields indexed by $a,b$ in a pseudoreal representation of the
gauge group determined by $T$, $A$ is an $SU(2)_R$ doublet index.
$\mu^\alpha_i$ are the hyperkahler moment maps, $\alpha =1,2,3$,
$i,j$ are gauge group indices, and $k$ is the matrix defined by the
Chern-Simons form. Therefore we have the equations
\begin{equation}\label{Flat}(k^{-1})^{i j} \mu_i^\alpha T^{a b}_j
q_b^A = 0,\end{equation}

There may be branches where $\mu^\alpha \neq 0$. On such branches,
the gauge group is not entirely Higgsed, since the moduli space
equation implies the matter fields are invariant under some gauge
transformations, determined by $k^{-1} \mu^\alpha$. Let's assume
for simplicity that the unbroken gauge group is Abelian. Hence we
may have a contribution to the moduli space from the dualized
gauge fields. The Chern-Simons coupling is a possible obstruction
to the dualization. Note that (\ref{Flat}) together with the
definition of the moment map,
$$\mu^\alpha_i = T^{a b}_i q^A_a q^B_b \Gamma^\alpha_{A B},$$ where $\Gamma^\alpha = \sigma^\alpha \sigma^2$
in terms of Pauli matrices are the tensor product coefficients,
implies that $$(k^{-1})^{i j} \mu^\alpha_i \mu^\beta_j = 0,$$ thus
$k^{-1} \mu$ lies in a subgroup, $H$, of the unbroken gauge group
which is null in the Chern-Simons form. This allows the
corresponding gauge fields to be dualized, and contribute to the
moduli space. At low energies, the ``extra'' directions in moduli
space allowed by $\mu^\alpha \neq 0$ should combine with the
dualized gauge bosons to give an hyperkahler manifold.

We can parameterize the extra directions in moduli space by the
expectation value of a monopole operator with magnetic flux in
$H$. When such a monopole is turned on, the Chern-Simons term is
not invariant under all constant gauge transformations, $\Lambda$,
due to the term $\frac{k}{4\pi} \Lambda \int_{S^2} F$. For a given
monopole background, the phase appearing in the partition
function, $e^{i S_{CS}}$, defines a map from the gauge group to
$U(1)$. One should only quotient by constant gauge transformations
that are in the kernel of this map for all allowed monopole
configurations (obeying the appropriate flux quantization) on a
given branch of the moduli space. Much like the Coulomb branch of
${\cal N}=4$ theories, the metric on these branches may receive
quantum corrections \cite{jafferis-yin}. Much information on the
Coulomb branch of ${\cal N}=4$ theories can be extracted by a
careful analysis of monopole operators \cite{Borokhov:2002cg},
\cite{Borokhov:2003yu}, \cite{Gaiotto:2008ak}. The same is true
for ${\cal N}=3$ CSM theories. We will present the analysis in the
next section

\section{Quantum corrected chiral ring of $\CN = 3$ CSM theories}

The moduli spaces and chiral ring of $\CN=3$ conformal field
theories display the rigidity of hyperk\"ahler manifolds. Indeed
chiral primary operators are the highest weight components of
$SU(2)_R$ multiplets, and their conformal dimension is determined
by the spin of the representation. The $\CN =2$ superpotential,
which determines the chiral ring, is fixed by $\CN=3$
supersymmetry, hence it is impossible for the F-term equations to
receive quantum corrections. Simple chiral primary operators made
out of the elementary scalar fields of the theory sit in $SU(2)_R$
multiplets determined immediately from the form of the operator,
hence their conformal dimension is unaffected by quantum
corrections.

If the expectation values of such simple chiral primary operators
were sufficient to parameterize the vacua of the theory, this would
be the end of the story. On the other hand, the $\CN=3$ CSM theories
we consider in this paper have a larger set of chiral primary
operators, and a more interesting moduli space. Indeed,as reviewed
in the previous section the moduli space parameterized by the $A,B$
bifundamental fields is unusually large due to the $\sum k_i=0$
constraint, which allows the moment maps of the gauge action to be
non-zero, and furthermore allows a certain combination of the gauge
fields to be dualized into an extra scalar field. Overall, the
moduli space has one extra hyperkahler dimension for each unbroken
$U(1)$, over which the shift of the dual photon acts
tri-holomorphically.

The simple chiral primary operators are insufficient to parameterize
the full moduli space, as they have charge zero under shifts of the
dual photon. Operators charged under the shift of the dual photons
have to carry magnetic charge, and can be realized as disorder
operators in the three dimensional field theory, as in
\cite{Borokhov:2002ib}, \cite{Borokhov:2002cg}

As we are dealing with a CFT, the definition is truly
straightforward: we can use the state-operator map, and simply look
at BPS states in the theory on a two-sphere, with magnetic flux on
the sphere. In a gauge theory with Yang-Mills coupling such states
are simply realized by turning on a constant gauge field on the
sphere in a specific $U(1)$ subgroup of the gauge group, and a
constant expectation value in the same $U(1)$ subgroup of one
adjoint scalar of the gauge multiplet. There are three scalars in
the gauge multiplet, and the choice of one of them corresponds to
the choice of an $N=2$ subalgebra. There are fermion zeromodes from
any fermion charged under the magnetic field, which need to be
properly quantized. The final result is that the BPS vacuum for the
fermion zeromodes carry an R-charge in the $N=2$ subalgebra equal to
\begin{equation}
Q=\frac{1}{2} (\sum_{i\in hyper} - \sum_{i \in vector}) |q_i|
\end{equation}

Here $q_i$ are the $U(1)$ gauge charges of fermions in either
hypermultiplets or vectormultiplets. If this R-charge is positive,
the states defined by different $\CN=2$ subalgebras can be organized
into a finite-dimensional $SU(2)_R$ multiplet of conformal dimension
$Q$. \footnote{If the R-charge is nonpositive, it signals a mistake
in the assumption that the theory flows to an IR fixed point with
the same $SU(2)_R$ R-symmetry as in the UV. \cite{Gaiotto:2008ak}}
To study monopoles in a CSM theory we can add a small Yang-Mills
coupling as a regulator of the theory. The main effect of the CS
coupling is that the Chern-Simons equations of motion are not
satisfied by a constant magnetic flux on the sphere in the absence
of an appropriate charge density, as $k *F = J$. Such a charge
density can be generated by acting on the naive vacuum with creation
operators of matter scalar fields in the s-wave. There is a certain
tension here: if a scalar field is charged under the $U(1)$ magnetic
field of the monopole, it has Landau levels on the sphere which do
not include an s-wave state, and have an energy greater than the
R-charge of the field. Hence only scalar fields with no charge under
the $U(1)$ subgroup used to define the monopole can be used to dress
the naive vacuum of the monopole. On the other hand, for generic
choices of the CS levels and monopole charges, the required $k *F$
charge has a component along this $U(1)$, and the CS equations of
motion can never be satisfied. This is why it is important that the
$U(1)$ charge should be null in the Chern-Simons form.

In the specific CSM theories we consider, with $\sum k_i=0$, a
monopole generated by a $U(1)$ embedded the same way (say by an
element $\bf{t}$ of the Lie algebra) in all gauge groups requires a
charge proportional to $(k_1 {\bf t},k_2 {\bf t},\cdots k_n {\bf
t})$, which is orthogonal to the $U(1)$ embedding $(\bf{t},
\bf{t},\cdots \bf{t})$. Moreover, such charge can be generated by
acting with creation operators from the bifundamental scalar fields,
which are not charged under the monopole $U(1)$. From now on we will
always consider such monopole operators. Overall, the dimension of
the monopole operator will be the sum of the two contributions
\begin{equation}
Q_0=\frac{1}{2} (\sum_{i\in hyper} - \sum_{i \in vector}) |q_i|
\end{equation}
and the dimension of the scalar fields used in the dressing.

Hence if we change the matter content of the theory, for example by
adding fundamental matter at some node, the extra charged
hypermultiplets will contribute to the dimension of the monopole
operator, and correct the dimensions of the chiral ring operators
charged under shifts of the dual photon, by an amount $\frac{1}{2} q
m$ proportional to the charge $q$ and to the number of fundamental
fields $m=m_1+m_2$. This exactly what we found in section
\ref{lift}.

\subsection{Examples of quantum corrected ``geometric'' moduli space}

We begin with the theory that arises at low energies from $N$ D3
branes intersecting an NS5, $(1,1)$ fivebrane, and D5 brane. The
classical moduli space simply consists of $\IC^4$, since the
vanishing of the bosonic potential implies that the fundamentals $a
= 0, \ b=0$. In \cite{ABJM}, the ring of chiral operators was found
to be $T A_i$, $\tilde T B_i$, where $T$ is the 't Hooft operator
that creates one unit of magnetic flux for the diagonal combination
of gauge fields (and $\tilde T$ creates $-1$ units of that magnetic
flux). That is, $\int_{S^2} F_+ = 1$. Note that this operator is
mutually local (ie. has a non-singular OPE) with the bifundamental
matter fields, since they are neutral under this combination of the
gauge groups.

In the $\CN = 6$ theory without the additional of the
fundamentals, these operators are dimension zero. We now argue
that the addition of the fundamental makes this operator have
dimension $1/2$, so the gauge invariant operators $T A_i$, $T'
B_i$ have dimension 1. Together with the dimension 1 mesonic
operators, $A^i B_j$, these form the 8 dimensional representation
of the flavor $SU(3)$, and exactly give the ring of chiral
operators on $T^* \IP^2$. There have been various proposals in the
literature for the conformal field theory dual to this AdS$_4
\times N^{0,1,0}$ \cite{Fabbri, Yee} that are significantly
different from ours, but we see here that the quantum correction
to the moduli space is crucial for finding the correct answer.

We can construct the dimension 1 currents of the quantum $SU(3)$
flavor symmetry, by acting with supersymmetry generators on the
dimension 1 operators, which are nothing else but the moment maps of
the flavor symmetry group.


\subsection{OPEs}

Consider a monopole operator, $T$, with magnetic flux determined by
a map $\rho: U(1) \rightarrow G$, up to gauge equivalence. As we
explained above, it may pick up an anomalous dimension, $q/2$, for
$q$ a positive integer. This operator lives in a dimension $q+1$
representation of the $SU(2)_R$. There is also a distinct conjugate
operator $\tilde T$, the 't Hooft operator with magnetic flux
associated to $\tilde\rho(e^{i\phi}) = 1/\rho(e^{i\phi})$, which has
the same anomalous dimension. Suppose we want to compute the OPE of
$T$ and $\tilde T$ in the chiral ring. Due to the Chern-Simons
terms, the Gauss' law in the monopole background is modified, so
some of the zero modes of the matter field must be excited on the
$S^2$. We want to focus on the contribution to the OPE from the
magnetic flux configuration itself; the results should be dressed
with matter operators appropriately in gauge invariant combinations.
As discussed in \cite{Borokhov:2002ib}\cite{Borokhov:2002cg}, one
should compute the partition function on a cylinder, $I \times S^1$,
with the monopole configuration on the $S^2$.

Suppose one considers a particular embedding of the $U(1)$ into the
gauge group. One should think of the monopole operator that would be
gauge invariant in the absence of Chern-Simons terms as being the
average over the group of the associated operator. In such a
particular configuration, one may apply the method of
\cite{Borokhov:2002ib}\cite{Borokhov:2002cg} to determine the OPE.
It is not difficult to convince oneself that only when the
configurations at the two ends of the cylinder are identical do any
contributions to the chiral ring appear. The result of this
calculation is that
$$T \tilde T \sim \int dg_G \ (\mu \cdot \operatorname{ad}_g(h))^q,$$ where $h \in \mathfrak{g}$
is the generator of the $U(1)$ embedding.

In the special case $G=U(1)$ we reproduce the result of section
\ref{lift}. A gauge invariant operator containing $T$ corresponds to
a function on moduli space which has a factor $u^q$. A gauge
invariant operator containing $\tilde T$ corresponds to a function
on moduli space which has a factor $v^q$. The OPE is expected to
reproduce the fact that $u^q v^q = (uv)^q$ can be rewritten as the
$q$-th power of the moment map of the baryonic symmetry $\sum p_i
\mu_i$ by the F-term relations .



\section{Branes and chiral operators}

Let us first identify the operators dual to gravitons in the
M-theory description. These will correspond to gravitons, D0
branes and their bound states in the IIA near horizon geometry.
Note that the dilaton is not constant in the internal manifold, so
the most natural definition of the ``pure'' D0 brane is the
operator of smallest dimension charged under the $U(1)_B$.

The simplest operators are the mesons $\Tr(A_i B_j)$ which are
neutral under both $U(1)$ isometries, and transform in the adjoint
of the $SU(2)$. These are dual to gravitons in M-theory with no
momentum along the $T^2$ isometry directions. Note that there are
fewer protected operators of this form than in the $\CN = 6$ theory,
since operators of the form $\Tr(C_I C^\dag_J)$ include the likes of
$\Tr(A_i A_j^\dag)$ which is unprotected in the $\CN=3$ theory. It
would be interesting to study these almost protected operators that
would fill out the $SU(4)_R$ multiplet in the ABJM theory with a
small number of fundamentals. On the field theory side, the
anomalous dimensions will arise from loops of fundamental fields. On
the gravity side, they are expected to arise from the interaction
with the D6 branes.

To construct operators charged under the $U(1)_F$, recall that the
bifundamental hypermultiplets are all charged equally. Thus the
operator with smallest dimension, $n/2$, is $\Tr(A_n A_{n-1} \dots
A_1)$. Chiral primaries constructed out of the matter fields alone
cannot receive quantum corrections in these $\CN=3$ theories.

The minimal monopole operator from which a chiral primary may be
constructed is $T$ with magnetic charge in a $U(1)$ subgroup of
the diagonal $U(N)$ in $U(N)^n$. It carries $k_i$ fundamental
indices under the i$^{th}$ gauge group, due to the Chern-Simons
coupling. Therefore one can form gauge invariant chiral operators
of the form $T \prod_i C_i^{d_i}$, for $d_i - d_{i+1} = k_i$, and
where by $C_i^{d_i}$ we mean $A_i^{d_i}$ if $d_i > 0$ and
$B_i^{-d_i}$ if $d_i < 0$. The obvious solution is to take $d_i =
q_i+ d$, the D5 charge of the i$^{th}$ fivebrane. This operator
will have dimension $\frac{1}{2}(\sum d_i + m)$, and charge $\sum
q_i k_i$ under the $U(1)_B$, so they are dual to D0 branes.

If there is more than one fundamental, the M-theory supergravity
description is never strictly valid, since the 3-Sasakian internal
manifold will have orbifold singularities. In the IIA description,
the near horizon geometry is a warped compactification, so AdS$_4$
curvature and string coupling depend on the position in the
internal six manifold. This limit will be valid when the curvature
and string coupling are small at the maximum value of the size of
the M-theory circle. As an estimate of that radius, we will use
the inverse of momentum of the lightest D0 brane, as determined
from the field theory analysis.

In particular, we find that $$R_{str}^2 \apgt
\frac{R_{M_7}^3}{m+k} = \frac{8\pi N^{1/2}}{\sqrt{m+2k}}.$$ The
radius of the 3-Sasakian manifold in eleven dimensionsal Planck
units is $R_M / \ell_P \sim N^{1/6} (m+k)^{1/3} (m+2k)^{-1/6}$.
Thus the size of the M-theory circle at its largest will be of
order $N^{1/6} (m+k)^{- 2/3} (m+2k)^{- 1/6}$. Therefore, IIA
supergravity will be valid when $$ m+2k \ll N \ll (m+k)^4
(m+2k).$$

Note that the field theory becomes weakly coupled even for fixed
Chern-Simons level if the number of fundamentals is much greater
than $N$. In the regime where there is a gravity dual, but $m >>
k$, these reproduce the results of \cite{Pelc} for the D2-D6
system in flat space. This is natural, since the geometry is
dominated by the $\IC^2 \times \IC^2/\IZ_m$ singularity near the
lift of the D6 branes.



\section{$\CN = 3$ Mass deformation}

In completely Higgsed branch of the moduli space, the effect of
turning on FI masses, which breaks conformal invariance, is to
modify the equations to
\begin{equation} \mu^\alpha_i = \zeta^\alpha_i ,\end{equation}
where $i$ indices the nodes in the quiver. This is the usual FI
deformation of the D6-D2 system. For generic $\zeta^\alpha_i$, the
geometric branch (with $A_i$, $B_i$ diagonal, and fundamentals set
to zero) will be lifted.

In the IIB picture, there are clearly $\CN=3$ mass deformations
corresponding to the relative positions of all of the fivebranes.
If there are no D5 branes, these precisely correspond to the FI
parameters of the CSM theory - the overall one does not change the
potential, one linear combination is non-geometric as seen in
\cite{Gomis}, and the rest correspond to (partial) hyperkahler
resolutions of the singularity. Once we have at least one
fundamental, the overall FI parameter gives it a mass,
\begin{equation} W = \sum_i \frac{1}{k_i}(A_i B_i - B_{i+1}
A_{i+1} +p_i^s q_i^s - \zeta_i)^2,
\end{equation} where $p_i^s$ are the fundamentals on the i$^{th}$
node. The FI deformation $\zeta_i = k_i \zeta$ completely cancels
except for giving masses to the fundamentals. In the case of a
single fundamental added to the ABJM theory, this corresponds to a
complete resolution of the hyperkahler singularity, $T^*
\IC\IP^2$. The OPE of the monopole operators gets modified to $T
\tilde T \sim \mu - \zeta$. This is precisely what one expects for
the hyperkahler resolution of the geometric branch, when it is not
lifted for $\sum \frac{\zeta_i}{k_i} = 0$.

It is also possible to give different masses to each fundamental,
while preserving $\CN=3 $ supersymmetry. This corresponds in the IIB
picture to separating the D5 branes. The completely Higgsed branch
of the moduli space will be lifted, generically, while the geometric
branch will have the $\IZ_m$ singularity resolved. The latter fact
can be seen by noting that the quantum correction to the moduli
space occurs along the locus where the fundamentals become massless.
Giving them different explicit mass terms, $m_i^s p_i^s q_i^s$,
means that this location will be different for each fundamental. In
particular, the fundamental, $p_i^s$, becomes massless when
$$A_i B_i - B_{i+1} A_{i+1} = k_i m_i^s.$$

\subsection*{Acknowledgments} We would like to thank J.~Maldacena, A.~Tomasiello and X.~Yin for discussions.
We thank S.~Franco, A.~Hanany and I.~Klebanov for pointing out a mistake in
an earlier version of the draft. D.G. is supported in part by the
DOE grant DE-FG02- 90ER40542 and in part by the Roger Dashen
membership in the Institute for Advanced Study. D.J. is supported
in part by DOE grant DE-FG02-96ER40959.


\begin{thebibliography}{}

\bibitem{ABJM}
  O.~Aharony, O.~Bergman, D.~L.~Jafferis and J.~Maldacena,
  ``N=6 superconformal Chern-Simons-matter theories, M2-branes and their gravity duals,''
  arXiv:0806.1218 [hep-th].

\bibitem{jafferis-yin}
  D.~L.~Jafferis and X.~Yin,
  ``Chern-Simons-Matter Theory and Mirror Symmetry,''
  arXiv:0810.1243 [hep-th].

\bibitem{imamura-kimura}
  Y.~Imamura and K.~Kimura,
  ``Coulomb branch of generalized ABJM models,''
  Prog.\ Theor.\ Phys.\  {\bf 120}, 509 (2008)
  [arXiv:0806.3727 [hep-th]].

\bibitem{jafferis-tomasiello}
  D.~L.~Jafferis and A.~Tomasiello,
  ``A simple class of N=3 gauge/gravity duals,''
  JHEP {\bf 0810}, 101 (2008)
  [arXiv:0808.0864 [hep-th]].

\bibitem{Hohenegger:2009as}
  S.~Hohenegger and I.~Kirsch,
  ``A note on the holography of Chern-Simons matter theories with flavour,''
  arXiv:0903.1730 [hep-th].

\bibitem{Gaiotto:2009mv}
  D.~Gaiotto and A.~Tomasiello,
  ``The gauge dual of Romans mass,''
  arXiv:0901.0969 [hep-th].

\bibitem{gaiotto-yin}
  D.~Gaiotto and X.~Yin,
  ``Notes on superconformal Chern-Simons-matter theories,''
  JHEP {\bf 0708}, 056 (2007)
  [arXiv:0704.3740 [hep-th]].

\bibitem{kitao-ohta-ohta}
  T.~Kitao, K.~Ohta and N.~Ohta,
  ``Three-dimensional gauge dynamics from brane configurations with (p,q)-fivebrane,''
  Nucl.\ Phys.\  B {\bf 539}, 79 (1999)
  [arXiv:hep-th/9808111].

\bibitem{bergman-hanany-karch-kol}
  O.~Bergman, A.~Hanany, A.~Karch and B.~Kol,
  ``Branes and supersymmetry breaking in 3D gauge theories,''
  JHEP {\bf 9910}, 036 (1999)
  [arXiv:hep-th/9908075].

\bibitem{GGPT}
  J.~P.~Gauntlett, G.~W.~Gibbons, G.~Papadopoulos and P.~K.~Townsend,
  ``Hyper-Kaehler manifolds and multiply intersecting branes,''
  Nucl.\ Phys.\  B {\bf 500}, 133 (1997)
  [arXiv:hep-th/9702202].

\bibitem{bielawski-dancer}
 R. Bielawski and A. Dancer,
 ``The geometry and topology of toric hyperkahler manifolds,''
  Comm. Anal. Geom. 8 (2000), no. 4, 727-760.

\bibitem{lee-yee}
  K.~M.~Lee and H.~U.~Yee,
  ``New $AdS_4\times X_7$ Geometries with $\CN=6$ in M Theory,''
  JHEP {\bf 0703}, 012 (2007)
  [arXiv:hep-th/0605214].

\bibitem{Borokhov:2002cg}
  V.~Borokhov, A.~Kapustin and X.~k.~Wu,
  ``Monopole operators and mirror symmetry in three dimensions,''
  JHEP {\bf 0212}, 044 (2002)
  [arXiv:hep-th/0207074].

\bibitem{Borokhov:2003yu}
  V.~Borokhov,
  ``Monopole operators in three-dimensional N = 4 SYM and mirror symmetry,''
  JHEP {\bf 0403}, 008 (2004)
  [arXiv:hep-th/0310254].

\bibitem{Gaiotto:2008ak}
  D.~Gaiotto and E.~Witten,
  ``S-Duality of Boundary Conditions In N=4 Super Yang-Mills Theory,''
  arXiv:0807.3720 [hep-th].

\bibitem{Borokhov:2002ib}
  V.~Borokhov, A.~Kapustin and X.~k.~Wu,
  ``Topological disorder operators in three-dimensional conformal field theory,''
  JHEP {\bf 0211}, 049 (2002)
  [arXiv:hep-th/0206054].

\bibitem{Fabbri}
  M.~Billo, D.~Fabbri, P.~Fre, P.~Merlatti and A.~Zaffaroni,
  ``Rings of short N = 3 superfields in three dimensions and M-theory on AdS(4) x N(0,1,0),''
  Class.\ Quant.\ Grav.\  {\bf 18}, 1269 (2001)
  [arXiv:hep-th/0005219].


\bibitem{Yee}
  H.~U.~Yee,
  ``AdS/CFT with tri-Sasakian manifolds,''
  Nucl.\ Phys.\  B {\bf 774}, 232 (2007)
  [arXiv:hep-th/0612002].

\bibitem{Pelc}
  O.~Pelc and R.~Siebelink,
  ``The D2-D6 system and a fibered AdS geometry,''
  Nucl.\ Phys.\  B {\bf 558}, 127 (1999)
  [arXiv:hep-th/9902045].


\bibitem{Gomis}
  J.~Gomis, D.~Rodriguez-Gomez, M.~Van Raamsdonk and H.~Verlinde,
  ``A Massive Study of M2-brane Proposals,''
  JHEP {\bf 0809}, 113 (2008)
  [arXiv:0807.1074 [hep-th]].


























\end{thebibliography}
\end{document}